----------
X-Sun-Data-Type: default
X-Sun-Data-Name: pl7.tex
X-Sun-Content-Lines: 458

\documentstyle[12pt]{article}
\newcommand{\pl}{\partial}
\begin{document}

\begin{center}
{\LARGE\bf q-oscillators, (non-)K$\ddot{\bf a}$hler manifolds
 and constrained dynamics}$^{ ^{ ^*}}$\\

\vskip 0.5cm
{\large\bf Sergey V. SHABANOV}$^{ ^{ ^{**}}}$

\vskip 0.5cm
{\em Service de Physique Th{\'e}orique de Saclay, CEA-Saclay\\
F-91191 Gif-sur-Yvette Cedex, FRANCE}
\end{center}

\begin{abstract}
It is shown that $q$-deformed quantum mechanics
($q$-deformed Heisenberg algebra)
can be interpreted as quantum mechanics on K$\ddot{\rm a}$hler manifolds, or
as a quantum theory with second (or first)-class constraints.
\end{abstract}

\vskip 0.2cm
{\bf 1}. In the present letter, a classical limit of multimode $q$-deformed
Heisenberg-Weyl algebras \cite{1},\cite{wor},
meaning $\hbar \rightarrow 0$ (rather
than $q\rightarrow 1$), is analyzed. As was show in \cite{2}, a
non-commutative phase space does not necessarily emerge
from $q$-deformed quantum mechanics in the classical
limit if one assumes $q$ to be a function of the
Plank constant and a certain
dimensional constant (its possible physical interpretation is discussed
in \cite{2}). In this approach, the
$q$-deformed Heisenberg-Weyl algebra produces a quadratic symplectic
structure in the classical limit. For example, the one-mode $q$-deformed
Heisenberg-Weyl algebra
\begin{equation}
\hat{b}\hat{b}^+-q^2\hat{b}^+\hat{b} =\hbar,
\end{equation}
where $\hat{b}$ and $\hat{b}^+$ are creation and destruction operators,
turns into the following symplectic structure \cite{2}
\begin{equation}
\{b,b^*\}=-i(1-b^*b/\beta )\ ,
\end{equation}
where $\beta $ is a constant and  $b$ and $b^*$ are commutative
complex coordinates on the phase space. One can easily be convinced that (2)
follows from (1) when taking the classical limit $[\ ,\ ]/i\hbar
\rightarrow \{\ ,\ \}$ as $\hbar \rightarrow 0$ and $\hat{b},
\ \hat{b}^+$ are changed by classical holomorphic variables $b,\ b^*$,
respectively. An assumption of the
existence of this limit actually yields $1-q^2=\hbar /\beta +O(\hbar
^2)$ because $[\hat{b},\hat{b}^+]/i\hbar =
-i(1-(1-q^2)\hat{b}\hat{b}^+/\hbar)$ in accordance with (1).
Notice that the formal limit $\hbar \rightarrow 0$ in the right-hand side
of (1) would lead to
a non-commutative phase space $bb^* =q^2b^*b$. Yet, a canonical
quantization of the symplectic structure (2) ($[\ ,\ ]=i\hbar \{\ ,\ \}$)
results in the "deformed" Heisenberg-Weyl algebra (1) \cite{2}, \cite{3}.

The modification of the symplectic structure (2) obviously leads to a
modification of Hamiltonian equations of motion. If the Hamiltonian is
assumed to be proportional to $b^*b$ (a harmonic $q$-oscillator), then
the frequency of oscillations becomes a function of the oscillator energy.
Thus, one can regard a harmonic $q$-oscillator as a familiar anharmonic
oscillator \cite{2}.

\noindent
\underline{\hspace*{8cm}}

\noindent
{\small $^*$Supported by an MRT grant of the government of France
and Russian Foundation of Fundamental Research, grant
93-02-3827.

\noindent
$ ^{**}$On leave from: {\em Laboratory of Theoretical Physics,
Joint Institute for Nuclear Research,\\
P.O.Box 79, Moscow, Russia}

\noindent
e-mail address: {\bf shabanov@amoco.saclay.cea.fr}}

\newpage
Known examples of the $q$-deformation of the Heisenberg algebra
($q$-particles) \cite{4}, \cite{5} can  also be treated as systems with
a "deformed" symplectic structure. Consider the following symplectic
structure
\begin{equation}
\{x,p\}=1-xp/\beta
\end{equation}
where $x$ and $p$ are the coordinate and momentum of a particle. A canonical
quantization of (3) gives $q$-deformed Heisenberg algebra discussed in
\cite{4}
\begin{equation}
\hat{p}\hat{x}-q\hat{x}\hat{p}=-i\hbar q^{1/2},\ \ \
\hat{p}^+=\hat{p},\ \ \ \hat{x}^+=\hat{x}\ ,
\end{equation}
with $q=e^{i\theta }$ and $\theta =\theta (\hbar /\beta )$ \cite{6}.
Indeed, by virtue of the canonical quantization rule, we have from (3)
$[\hat{x},\hat{p}]=i\hbar(1-(\hat{x}\hat{p}+\hat{p}\hat{x})/2\beta)$.
Relation (4) is obtained then by renormalizing the operators $\hat{p}$
and $\hat{x}$ with the coefficient $|1-i\hbar/2\beta|^{1/2}$, and
$q=(1+i\hbar/2\beta)/(1-i\hbar/2\beta)$.

The
Hamiltonian equation of motion $\dot{x}=\{x,H\}$ of a free particle
$H=p^2/2$ induced by (3) coincides with the equation of motion
for a particle with friction.
The friction coefficient depends on the deformation parameter $\beta $
and the particle generalized momentum $p$. Notice that $\dot{p}=\{p,H\}
=0$ therefore $p=const$.

A lattice quantum mechanics \cite{5},
\cite{7} appearing upon a deformation of the
Heisenberg algebra with a real $q$ \cite{5}
\begin{equation}
\hat{p}\hat{x}-q\hat{x}\hat{p}=-i\hbar,\ \
\hat{x}\hat{p}^+-q\hat{p}^+\hat{x}=i\hbar,\ \
\hat{p}^+\hat{p}=q\hat{p}\hat{p}^+,\ \ \hat{x}^+=\hat{x}
\end{equation}
can be obtained by quantizing a degenerate symplectic structure
\begin{equation}
\{x,p\}=1-ixp/\beta ,\ \ \{x,p^*\}=1+ixp^*/\beta ,\ \
\{p^*,p\}=ipp^*/\beta \ .
\end{equation}
The degeneracy ia associated with the existence of an absolute integral of
motion
\begin{equation}
C=pp^*x/\beta -i(p-p^*)
\end{equation}
which commutes with all symplectic coordinates
$\{C,x\}=\{C,p\}=\{C,p^*\}=0$. Therefore, the system never leaves the
surface $C={\rm const}$ in due course. A phase space of the system is a
two-dimensional surface $C={\rm const}$. In quantum theory, eigen values
of the Casimir operator $\hat{C}$ determine irreducible representation of
the algebra (5) \cite{5}.

A straightforward application of the canonical quantization rule to (6)
meets the operator ordering ambiguity in the right-hand side of the
commutation relation $[,]=i\hbar\{,\}$ (see also a Remark in p.3).
The operator ordering should be chosen so that
the Jacobi identity is fulfilled on the quantum level
\cite{Berezin}.
It is remarkable that any operator ordering consistent with the Jacobi
identity results in the algebra (5). Different choices of the operator
ordering correspond to variations of terms $O(\hbar^2)$ in
$q=q(\hbar,\beta)$ \cite{2}.

To get (5) from (6), one can, for instance, postulate the first commutation
relation as follows $[\hat{x},\hat{p}]=i\hbar(1-i\hat{x}\hat{p}/\beta)$,
then $[\hat{x},\hat{p}^+]$ is obtained by the Hermitian conjugation
of the first one, assuming $\hat{x}=\hat{x}^+$, while the operator
ordering in the last commutation relation in (6) is fixed by the Jacobi
identity and reads $[\hat{p}^+,\hat{p}]=-\hbar\hat{p}\hat{p}^+/\beta$. So,
$q=1-\hbar/\beta$.

So, the $q$-deformed Heisenberg algebra can appear as
a result of quantizing a quadratic symplectic structure
\begin{equation}
\{\theta ^j,\theta ^k\}
=\stackrel{\circ}\omega^{jk}+c^{jk}_{in}\theta ^i\theta ^n
\end{equation}
where $\theta ^j$ is a set of real phase-space coordinates,
$\stackrel{\circ}\omega^{jk}$ is the canonical
symplectic structure and $c^{jk}_{in}$
are "deformation" constants chosen so that the Jacobi identity for (8) is
satisfied.

Below we shall demonstrate that the symplectic structure associated with
the $SU_q(n)$-covariant deformation of the Heisenberg-Weyl algebra is
related to quantum mechanics on K$\ddot{\rm a}$hler manifolds. We shall also
show
that physical phase-space variables in constrained quantum mechanics
may naturally form a $q$-deformed Heisenberg-Weyl algebra.

{\bf 2}. The following $q$-deformed commutation relations remain untouched
under the action of the quantum group $SU_q(n)$ \cite{wor}
\begin{eqnarray}
\hat{a}_i\hat{a}_j &= &q\hat{a}_j\hat{a}_i,\ \ \hat{a}_i^+\hat{a}_j^+=
\frac{1}{q}\hat{a}^+_j\hat{a}^+_i,\ \ i<j\\
\hat{a}_i\hat{a}_j^+&= &q\hat{a}^+_j\hat{a}_i,\ \ i\neq j\\
\hat{a}_i\hat{a}^+_i&-& q^2\hat{a}^+_i\hat{a}_i=\hbar
+(q^2-1)\sum\limits_{k<i}^{} \hat{a}_k^+\hat{a}_k\ .
\end{eqnarray}
To obtain a corresponding symplectic structure in a classical theory, one
may use the rule $[\ ,\ ]/i\hbar \rightarrow\{\ ,\ \}$ as $\hbar
\rightarrow 0$ and $\hat{a}_i,\ \hat{a}_j^+$ are simultaneously to be changed
by classical holomorphic variables $a_i,\ a_j^*$.
However, this is just a formal rule which sometimes helps
to guess a correct classical limit of a given quantum theory (see a
rigorous consideration of the classical limit in \cite{Berezin}).
To ensure that this rule works
for the algebra (9)-(11), we notice that by means of a
transformation proposed in \cite{cha} the commutation relations (9)-(11)
can be "diagonalized", be transformed to the form (1) for each oscillator
mode, while operators of different modes commute amongst each other. For
the commutation relation (1), the validity of the rule $[\ ,\ ]/i\hbar
\rightarrow \{\ ,\ \}$ can be rigorously established
in the framework of the path
integral formalism \cite{2}. Therefore the above mentioned formal approach
gives a correct classical mechanics in our case.
Assuming again $1-q=\hbar /\beta +O(\hbar ^2)$
(otherwise there is no commutative phase space in the classical theory) we
arrive at the following Poisson bracket structure
\begin{eqnarray}
\{a_k,a_j\} &= &ia_ka_j/\beta ,\ \ \{a^*_k,a^*_j\}=-ia^*_ka^*_j/\beta ,
\ \ k<j\\
\{a_k,a^*_j\} &= &ia_ka^*_j/\beta ,\ \ \ k\neq j\ ;\\
\{a_j,a^*_j\}&=&-i\left(1-\frac{2}{\beta}\sum\limits_{k=1}^{j}
a^*_ka_k\right )\ ,
\end{eqnarray}
where $a_j,a^*_j$ are holomorphic coordinates on a phase space.

Let us recall now a basic definition of the K$\ddot{\rm a}$hler manifold
\cite{Berezin}.
Let $z^i$ and $z^{k*}$ are complex coordinates on a manifold ${\cal M}$
and $g_{i\bar{k}}(z,z^*)$ is a metric tensor on it such that the
interval on ${\cal M}$ has the form $ds^2=g_{i\bar{k}}dz^idz^{k*}$ and
\begin{equation}
g_{i\bar{k}}=\pl ^2\phi /\pl z^i\pl z^{k*}\ ;
\end{equation}
the scalar function $\phi $ is called the K$\ddot{\rm a}$hler potential, and
${\cal M}$
is called the K$\ddot{\rm a}$hler manifold. A K$\ddot{\rm a}$hler manifold
turns into a symplectic
manifold if the following symplectic structure is introduced on it
\begin{equation}
\{A,B\}=-ig^{\bar{j}k}\left(\frac{\pl A}{\pl z^k}\frac{\pl B}{\pl
z^{j*}}-\frac{\pl A}{\pl z^{j*}}\frac{\pl B}{\pl z^k}\right)
\end{equation}
for any two functions $A$ and $B$ of $z,\ z^*$, where $g^{\bar{k}i}$
is a matrix inverse to (15). The Poisson bracket thus defined obey the
Jacobi identity due to the property (15) \cite{Berezin}.

It is easy to see that the Poisson brackets (12)-(14) are not of the
K$\ddot{\rm a}$hlerian type because of (12). However, they can be transformed
to
the form (16).
Indeed, the algebra (12)-(14) admits the following representation
\begin{equation}
a_i=z^i\prod\limits_{k=1}^{i-1}(1-2z^kz^{k*}/\beta )^{1/2}
\end{equation}
and $a^*_i$ is obtained by a complex conjugation of (17), where
\begin{equation}
\{z^j,z^{k*}\}=-i(1-2z^jz^{j*}/\beta )\delta ^{j\bar{k}}
\end{equation}
and $\{z^j,z^k\}=\{z^{k*},z^{j*}\}=0$.
Therefore, the K$\ddot{\rm a}$hler metric reads
\begin{equation}
g_{i\bar {k}}=\delta _{i\bar{k}}(1-2z^kz^{k*}/\beta )^{-1}\ .
\end{equation}
Representing the K$\ddot{\rm a}$hler potential in the form
\begin{equation}
\phi =\frac{\beta}{2}\sum\limits_{i}^{}\varphi \left( \frac{2z^iz^{i*}}{\beta}
\right)
\end{equation}
and substituting (20) and (19) into (15) we obtain
\begin{equation}
\varphi (x)=Li_2(x)=\sum\limits_{k=1}^{\infty}\frac{x^k}{k^2}\ ,\ \
|x|<1\ ,
\end{equation}
with $Li_2$ being the Euler dilogarithm.

So, the $SU_q(n)-$covariant deformation
of the Heisenberg-Weyl algebra describes
a quantum theory on a K$\ddot{\rm a}$hler manifold with the potential (20),
(21).

The phase-space manifold with the metric (19) is curved. The scalar
curvature corresponding to the metric (19),
\begin{equation}
R=\sum_i\frac{8}{\beta}\left(1-\frac{2z^iz^{i*}}{\beta}\right)^{-1}\ ,
\end{equation}
tends to infinity as any of variables $z^i$ approaches the
circle $|z^i|^2=\beta/2$, assuming $\beta >0$. Therefore, for
positive $\beta$ the phase space turns out to be compact, while
for negative $\beta$ the function (22) is regular on the entire
complex plane.

{\bf 3}. Another more general "interpretation" of the $q$-deformation of the
Heisenberg-Weyl algebra can be achieved in the framework of the
constrained dynamics. It is known since long time ago that a non-trivial
symplectic structure may occur through second-class constraints
\cite{Dir} in dynamical systems.
Let $\varphi _a(\xi)=0,\ \ a=1,2,\ldots ,2M$, are second-class
constraints on a phase space spanned by coordinates $\xi^i$,
i.e. the matrix $\{\varphi _a,\varphi _b\}=\Delta _{ab}$ is not
degenerate, where $\{\xi ^i,\xi ^j\}=\stackrel{\circ}\omega ^{ij}$ is the
canonical symplectic structure. Let $\xi ^i=\xi ^i(\theta )$ be a solution
of constraints where physical variables $\theta ^\alpha ,\ \ \alpha
=1,2,\ldots ,2(N-M)$ are coordinates of a physical phase space,
$\varphi_a(\xi(\theta))\equiv 0$. Then the
symplectic structure on the physical phase space is induced by the Dirac
bracket \cite{Dir}
\begin{equation}
\{A,B\}_D=\{A,B\}-\{A,\varphi _a\}\Delta^{ab}\{\varphi _b,B\}
\end{equation}
projected on the surface $\xi ^i=\xi ^i(\theta )$, here
$\Delta^{ab}\Delta_{bc}=\delta ^a_c$.

The induced symplectic structure might not coincide with the canonical
one, i.e. it might turn out to be "deformed". Therefore, one can ask the
question: is it possible to find such second-class constrained system
that the symplectic structure induced by the Dirac bracket on the physical
phase space has the quadratic form (8)? The answer is positive for the
simplest $q$-deformed systems considered in p.1 \cite{6}. A generalization
is rather trivial.

Let $\omega ^{ij}(\theta )$ be a non-constant
symplectic structure and $\omega
_{ij}\omega ^{jk}=\delta ^k_j$. Let us
extend the initial phase space spanned
by $\theta ^j$ by adding new variables $\pi _j$ and postulate the canonical
symplectic structure on the extended phase space, $\{\theta ^j,\theta
^k\}=\{\pi _i,\pi _k\}=0$ and $\{\theta ^j\pi _k\}=\delta ^j_k$, i.e. the
initial phase space is a configuration space in the extended theory.
Following \cite{Bat} we introduce second class constraints as
\begin{equation}
\varphi _i(\pi ,\theta )=\pi _i+\bar{\omega }_{ij}(\theta )\theta ^j=0
\end{equation}
where
\begin{equation}
\bar{\omega }_{ij}(\theta )=(\theta ^i\frac{\pl}{\pl \theta
^i}+2)^{-1}\omega _{ij}(\theta )=\int\limits_{0}^{1} d\alpha \alpha\omega
_{ij}(\alpha \theta ) \ .
\end{equation}
Then \cite{Bat}
\begin{equation}
\{\theta ^i,\theta ^j\}_D=-\{\theta ^i\varphi _k\}\Delta ^{kn}\{\varphi
_n,\theta ^j\}=\omega ^{ij}(\theta )\ .
\end{equation}
For the symplectic structures (18) or (12)-(14), the integral (25) can be
taken explicitly. Thus, $q$-deformed quantum mechanics can appear upon a
quantization of second-class constrained systems.

{\em Remark}. Quantization of the quadratic symplectic
structure (8) (induced by the Dirac bracket (23)) is not obvious because
of the operation ordering. A naive application of the formal rule $[\ ,\
]=i\hbar \{\ ,\ \}$ can break down the Jacobi identity in the quantum
theory (or the associativity of the quantum algebra). One way to obtain an
associative quantum theory is to assume that the coefficients
$c^{kn}_{ij}$ are also to be changed
upon quantizing, $c^{kn}_{ij}\rightarrow
\tilde{c}^{kn}_{ij}(\hbar )$ so that $\tilde{c}^{kn}_{ij}(\hbar
=0)=c^{kn}_{ij}$ and $\tilde{c}^{kn}_{ij}$ provide the associativity of
the quantum theory. The latter yields some algebraic equations for
$\tilde{c}^{kn}_{ij}$ of the Yang-Baxter-Hecke type to be solved.

Another way to manage the operator ordering problem is to convert the
second-class constrained dynamics (24) into the first-class ones with
a sequent quantization \cite{Fad}, \cite{Bat}
$\footnote{This procedure is equivalent to a quantization via the
Darboux variables for a given symplectic form. However, the goal
of the conversion method is that it allows us to avoid an explicit
construction of the Darboux variables (which is a rather hard problem
in general).}$.
A curious observation in
this approach is that $q$-deformed commutation relations can also
occur through reducing a first-class constrained (gauge) quantum system to
physical (gauge-invariant) variables. One should point out that
a quadratic symplectic
structure is just a particular case in the framework of the generalized
canonical quantization of curved phase spaces developed in \cite{Bat}.

{\bf 4}. For any symplectic matrix $\omega _{ij}(\theta )$ obeying the
Jacobi identity $\pl _k\omega _{ij}+{\rm cycle}(k,i,j)=0$, there exist
local Darboux coordinates in which the symplectic structure has the
canonical form \cite{Arnold}. The Darboux variables for the Poisson bracket
(18) and, hence, for (12)-(14) (due to the relation (17))
can be explicitly found \cite{2}.
Therefore, the quadratic "deformation" locally looks like a special
non-canonical transformation of the ordinary (Darboux) phase-space
coordinates \cite{2}. From the mathematical point of view, all phase-space
coordinate systems are to be treated on equal footing. In a physical
theory, phase-space coordinates are associated with observables, which
makes some particular coordinates to be dynamically distinguished. For
example, the excitations of various physical systems can be modelled
through $q$-oscillators \cite{mas}, which means that all complicated
interactions in a physical system can be accumulated into $q$-deformed
commutation relations. So, from this point of view the
$q$-deformed quantum mechanics appears to be
an effective theory for describing physical excitations. It has
been, actually, illustrated in p.1 with examples of
the harmonic $q$-oscillator and $q$-particle (which are dynamically
equivalent to an anharmonic oscillator and a particle with friction,
respectively).

In contrast with the above said, the interpretation within constrained
dynamics does not imply, in general, any non-trivial interaction leading
to "$q$-deformed" excitations. The $q$-deformation of the algebra of
observables appears kinematically upon eliminating all unphysical
(gauge) degrees of freedom (i.e. after solving constraints).
\begin{center}
{\bf Acknowledgement}
\end{center}
The author is kindly grateful to Prof. I.A. Batalin for useful
discussions.

\end{document}